\documentclass[12pt]{iopart}  
\usepackage[utf8]{inputenc}

\usepackage{iopams}      
\usepackage{graphicx}
\usepackage{xcolor}
\usepackage{float}
\usepackage{caption}
\usepackage{subcaption}
\usepackage{adjustbox}
\usepackage{booktabs}
\usepackage{colortbl}
\usepackage{hyperref}
\newcommand{\doi}[1]{\href{https://doi.org/#1}{doi:#1}}
\newcommand{\arxiv}[1]{\href{https://arxiv.org/abs/#1}{arXiv:#1}}
\usepackage{tikz}
\usepackage{pgfplots}
\pgfplotsset{compat=newest}
\usetikzlibrary{arrows.meta,positioning,shapes.geometric,plotmarks,calc,pgfplots.groupplots,decorations.pathmorphing,decorations.pathreplacing, positioning}

\usepackage[margin=0.85in]{geometry}

\definecolor{costlow}{RGB}{198,239,206}
\definecolor{costmid}{RGB}{255,235,156}
\definecolor{costhigh}{RGB}{255,199,206}
\definecolor{yes}{RGB}{184,225,134}
\definecolor{no}{RGB}{255,199,206}

\def\begred{\begin{color}{red}}
\def\endred{\end{color}\ }
\def\begblue{\begin{color}{blue}}
\def\begorange{\begin{color}{orange}}
\def\begapricot{\begin{color}{apricot}}
\def\endapricot{\end{color}\ }
\def\endorange{\end{color}\ }
\def\endblue{\end{color}\ }
\def\beggreen{\begin{color}{green}}
\def\endgreen{\end{color}\ }
\def\begbrown{\begin{color}{brown}}
\def\endbrown{\end{color}\ }
\def\begblack{\begin{color}{black}}
\def\begcyan{\begin{color}{cyan}}
\def\endcyan{\end{color}\ }

\renewcommand{\thesection}{\Roman{section}}
\renewcommand{\thesubsection}{\Alph{subsection}}
\usepackage{ifthen}
\usepackage{xparse}
\usepackage{comment}
\usepackage{verbatim}
\usepackage{etoolbox}

\newtoggle{dszmode}

\newcommand{\DSZ}{%
  \iftoggle{dszmode}{%
    \par\vspace{0.5em}%
    \begingroup\color{red}\textbf{DSZ: work to be done}\endgroup%
    \vspace{0.5em}\par%
  }{}%
}

\begin{document}

\title{Entangled Quantum Negative Energy Teleportation as a Probe of Semiclassical Gravity}

\author{Daniel S. Zachary$^{1}$}


\address{$^{1}$ Advanced Academic Program, Krieger School of Arts and Sciences, Johns Hopkins University, Washington DC,  555 Pennsylvania Avenue, NW, 20001, USA}

\ead{d.s.zachary@jhu.edu}


\vspace{10pt}
\begin{indented}
\item[]\today
\end{indented}

\begin{abstract}
We investigate the generation of semiclassical spacetime curvature via localized negative energy densities created by quantum energy teleportation (QET) and Casimir-enhanced confinement. Using realistic noise models and experimental architectures, we compute signal-to-noise ratios for detecting the resulting Ricci curvature via atomic clocks, interferometry, and optomechanical strain readout. We propose synchronization and squeezing strategies to enhance detectability and simulate spatial curvature profiles from focused QET pulses. Finally, we introduce a speculative framework---the Quantum-Curvature Compression Channel---as an experimentally motivated alternative to warp-drive geometries, enabling apparent geodesic compression through synchronized quantum energy operations. Our results clarify the experimental path toward laboratory tests of exotic stress-energy and semiclassical gravity effects.
\end{abstract}



\maketitle


\section{Introduction}\label{sec:Intro}

The interface between quantum field theory and general relativity remains one of the most profound and unresolved frontiers in theoretical physics. Among its open questions, a particularly intriguing challenge is understanding how quantum energy fluctuations—especially negative energy densities—contribute to spacetime curvature within the semiclassical regime \cite{DeWitt1979,Ford1995}.

Negative energy is permitted in quantum theory, manifesting in squeezed states, the Casimir effect, and vacuum entanglement \cite{FordRoman1990, FewsterRoman2005, Hotta2008}. Yet despite decades of theoretical exploration, its gravitational consequences remain experimentally unverified. In semiclassical gravity, the expectation value of the stress-energy tensor acts as a source term in the Einstein equations. However, no controlled laboratory experiment has demonstrated a measurable gravitational response to an engineered quantum state.

In this work, we propose and model a tabletop experimental platform designed to generate and detect the gravitational effects of negative energy via quantum energy teleportation (QET) \cite{Hotta2008}. By coupling entangled quantum systems to local field observables, QET enables spatially separated energy extraction that relies on quantum correlations rather than direct energy transfer. We analyze how such operations can produce localized negative energy densities, and how these in turn can induce weak but detectable curvature signatures.

The proposed framework—termed QIX (Quantum Interferometric Exchange)—integrates QET with high-precision metrology, including optical interferometers and atomic clocks \cite{Kolkowitz2016}, to probe curvature perturbations arising from quantum field operations. In addition to testing core predictions of semiclassical gravity, this platform offers a laboratory venue for exploring deeper conjectures such as ER=EPR \cite{Maldacena2013}, which posit that entanglement may have a direct geometric realization.

This study builds upon and extends theoretical models introduced in \cite{Zachary2025}, incorporating new signal-to-noise analyses and experimental designs.

\medskip

The remainder of this paper is organized as follows. Section~\ref{sec:theory} introduces the theoretical framework of QET and outlines the semiclassical response to engineered quantum energy distributions. Section~\ref{sec:ExArch_3} presents the QIX architecture, including candidate coupling schemes and sensing platforms. Sections~\ref{sec:FocNegEnergy_4} and~\ref{sec:ObsSigSen_5} examine methods for concentrating negative energy, simulating the resulting curvature, and evaluating detector response. Section~\ref{sec:ConNegEnSup_6} provides a comparative sensitivity analysis across various quantum sensors. Section~\ref{sec:QIXC_8} provides a 
a speculative extension to this work and 
Section~\ref{sec:conc_9} concludes with implications for future experiments and prospects for testing semiclassical gravity using entangled quantum systems.


\section{Theoretical Framework}
\label{sec:theory}

Quantum Energy Teleportation (QET) is a protocol that exploits spatial entanglement in quantum fields to extract energy at one location through local operations conditioned on measurements performed elsewhere \cite{Hotta2008,Hotta2009}. Unlike conventional teleportation of quantum states, QET enables the transfer of energy without violating locality or energy conservation, owing to its reliance on pre-existing field correlations and the injection of negative energy in the intermediate region.

The QET protocol typically involves two agents, Alice and Bob, coupled locally to a quantum field. Alice performs a measurement on her local field observables, injecting energy into the system while projecting the field into a conditionally squeezed state. She then transmits her classical measurement outcome to Bob, who performs a unitary operation on his local field observables. Under suitable conditions, Bob can extract energy from the field, even though the field is initially in its vacuum state. The energy gained by Bob is ultimately drawn from the correlations that existed between the two field regions, not from any energy transmitted between them.
\DSZ 

This process necessarily creates localized regions of negative energy density, as required by quantum inequalities \cite{FordRoman1997}. These negative energy densities are transient and bounded but play a key role in maintaining consistency with quantum field theory. Of particular interest is the stress-energy tensor expectation value $\langle T_{\mu\nu}(x) \rangle$ during and after the QET process. In semiclassical gravity, this tensor acts as the source of curvature via the semiclassical Einstein equation,

\begin{equation}
    G_{\mu\nu}(x) = 8\pi G \langle T_{\mu\nu}(x) \rangle.
    \label{eq:Einstein}
\end{equation}

Although the magnitude of $\langle T_{\mu\nu}(x) \rangle$ produced in a tabletop QET setup is small, the corresponding curvature perturbations are in principle observable with sufficiently sensitive probes. Because these effects depend on nonlocal field correlations and entanglement structure, their gravitational consequences may provide insights into whether semiclassical gravity fully accounts for quantum coherence, or whether more fundamental modifications are required.

In this work, we focus on a class of QET protocols in which the field is modeled as a scalar vacuum, and the local interactions are mediated by Unruh–DeWitt–type detectors \cite{DeWitt1979}. We construct the effective stress-energy tensor induced by the protocol and simulate the resulting spacetime curvature within a linearized gravity approximation. Our goal is to determine whether this curvature, particularly its time-resolved spatial profile, is detectable via interferometric or atomic clock shifts.

These QET-induced effects represent an experimentally accessible regime of semiclassical gravity, enabling a controlled study of quantum-coherent contributions to the gravitational field. The interplay between entanglement, negative energy, and curvature response opens a potential pathway toward resolving the semiclassical limit of quantum gravity.


\section{Experimental Architecture}
\label{sec:ExArch_3}

To observe curvature signatures arising from quantum energy teleportation (QET), we propose a tabletop experimental platform referred to as QIX (Quantum Interferometric Exchange). This setup integrates quantum field couplings with entanglement resources and high-precision metrological devices, designed to detect weak curvature perturbations generated by localized negative energy densities.

The QIX system is composed of four essential subsystems. The first is a source of entanglement between spatially separated quantum systems, such as field modes, superconducting qubits, or cold atomic ensembles. These serve as the quantum resource necessary for the QET protocol. The second component involves local field couplers modeled as Unruh–DeWitt detectors. These detectors are effectively two-level systems that interact locally with a quantum field via time-dependent switching and spatial smearing functions. The interaction Hamiltonian takes the form

\begin{equation}
H_{\mathrm{int}}(t) = \lambda(t)\, \chi(\mathbf{x})\, \hat{\mu}(t)\, \hat{\phi}(t, \mathbf{x})
\label{eq:Hint}
\end{equation}


where $\lambda(t)$ describes the temporal switching, $\chi(\mathbf{x})$ specifies the spatial profile of the interaction, $\hat{\mu}(t)$ is the detector monopole operator, and $\hat{\phi}(\mathbf{x},t)$ is the scalar field operator. This interaction enables Alice to inject energy into the field and Bob to extract it, resulting in a localized redistribution of the field's stress-energy density.

The third subsystem is a classical communication channel that transmits Alice’s measurement result to Bob, enabling him to apply a conditional operation without violating causality or energy conservation. Importantly, the classical channel does not transmit energy. The fourth and final subsystem consists of curvature-sensitive detectors positioned in regions where QET is predicted to induce localized negative energy. These detectors include optical interferometers, atomic clocks, and compliant resonators, each sensitive to different aspects of the perturbed metric—such as proper time shifts, phase delays, or redshift anomalies.

All subsystems must be precisely synchronized in time and space. The detector couplings are activated according to a predefined sequence, with control over gate timing, spatial separation, and coupling strength. This ensures that the expected stress-energy perturbation is spatially localized and temporally sharp, allowing differential measurements before and after the QET operation. The time-resolved curvature signal can then be extracted using matched filtering or phase-shift reconstruction techniques, depending on the detector employed.

Realizations of the QIX platform can be tailored to various experimental domains. Optical implementations may use squeezed states in high-finesse cavities to achieve sub-vacuum energy densities. Superconducting qubit systems can provide tunable local interactions with coplanar waveguides or resonators in cryogenic environments. Cold atom clocks, known for their extreme timing stability, offer another pathway to observe small shifts in proper time due to transient curvature.


\begin{figure}[h]
\centering
\begin{tikzpicture}[scale=1.5]

  \draw[->] (-0.5, 0) -- (4.5, 0) node[right] {Space \(x\)};
  \draw[->] (0, -0.5) -- (0, 4.5) node[above] {Time \(t\)};

  \coordinate (A) at (0,0);      
  \coordinate (B) at (3,3);      

  \filldraw[blue] (A) circle (1.5pt) node[below left] {Alice};
  \node[blue] at (-0.5,0.3) {$\hat{A}(x,t)$};

  \filldraw[red] (B) circle (1.5pt) node[above right] {Bob};
  \node[red] at (3.5,2.7) {$\hat{B}(x',t')$};

  \draw[->, thick, dashed, gray] (A) -- (B) node[midway, below right] {signal};

  \coordinate (L) at (0,1.5);  
  \coordinate (R) at (3,1.5);  

  \draw[gray!60, fill=gray!10] (A) -- (L) -- (B) -- (R) -- cycle;

  \draw[dotted] (A) -- (L);
  \draw[dotted] (A) -- (R);
  \draw[dotted] (B) -- (L);
  \draw[dotted] (B) -- (R);

  \node[gray!70] at (1.5,1.5) {\small Causal Field Region};

\end{tikzpicture}

\caption{Schematic of the QIX architecture. Alice and Bob are locally coupled to a quantum field via Unruh–DeWitt detectors. 
A classical signal from Alice enables Bob to perform a conditional operation that extracts energy, producing a region of negative energy density. A curvature-sensitive detector is positioned nearby to measure the induced metric perturbation.}
\label{fig:QIXsetup}
\end{figure}
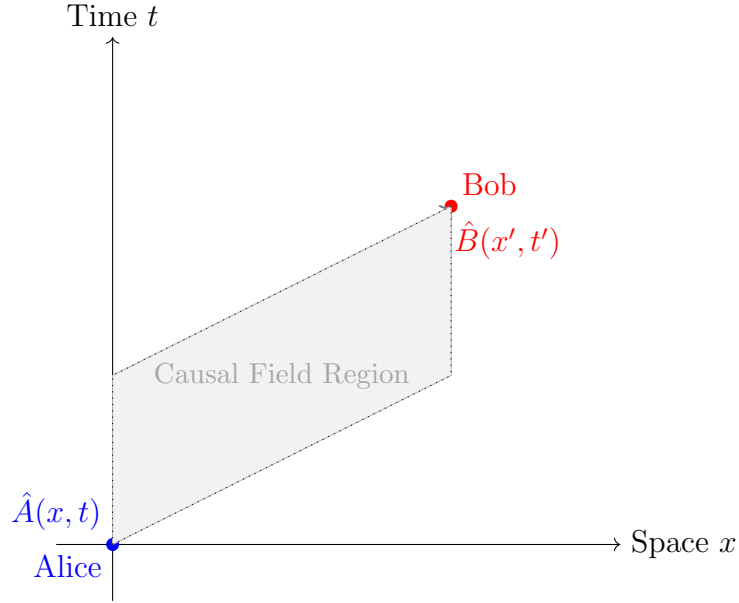

This integrated architecture enables QET protocols to be implemented within realistic laboratory settings, with the possibility of detecting curvature responses to quantum-coherent operations. In the following sections, we model the negative energy profile generated by these couplings and simulate the resulting spacetime curvature and observable signals.


\section{Negative Energy Concentration and Curvature Modeling}
\label{sec:FocNegEnergy_4}

The QET protocol generates localized negative energy density as a result of the nonlocal correlations in the quantum field exploited by sequential measurements and conditional operations. These localized negative regions arise between the source (Alice) and the receiver (Bob), and are essential to maintaining energy conservation in the field while allowing Bob to extract net positive energy.

To characterize this effect, we consider the expectation value of the renormalized stress-energy tensor $\langle T_{\mu\nu}(x) \rangle$ during the QET sequence. For a scalar field in Minkowski spacetime initially prepared in the vacuum state, local interaction via the Unruh–DeWitt coupling modifies the field state nontrivially. 
\DSZ 
When Alice performs her measurement and communicates the result to Bob, his local operation modifies the field state further, inducing regions in which $\langle T_{00}(x) \rangle < 0$ for a finite duration.

The spatial and temporal profile of this negative energy region depends on the detector parameters, including the switching function $\lambda(t)$, the spatial smearing profile $\chi(\mathbf{x})$, the separation $d$ between Alice and Bob, and the timing offset between their operations. We approximate the resulting energy distribution using a superposition of Gaussian functions centered between the interaction points, with width determined by the detector smearing scale $\sigma$ and amplitude related to the QET coupling strength. A representative form is:
\begin{equation}
    \langle T_{00}(x,t) \rangle \approx -\epsilon\, \exp\left(-\frac{(x - x_0)^2}{2\sigma^2} - \frac{(t - t_0)^2}{2\tau^2} \right),
    \label{eq:NegEnergyGaussian}
\end{equation}
where $\epsilon$ encodes the effective energy density magnitude, and $(x_0, t_0)$ denotes the center of the negative energy pulse.

To evaluate the gravitational response, we solve the semiclassical Einstein equations in the weak-field regime. Assuming perturbations around flat spacetime and working in harmonic gauge, the linearized Einstein tensor $G_{\mu\nu}$ is proportional to $\langle T_{\mu\nu} \rangle$:

\begin{equation}
    \Box \bar{h}_{\mu\nu}(x) = -16\pi G\, \langle T_{\mu\nu}(x) \rangle,
    \label{eq:LinearEinstein}
\end{equation}
\DSZ 

where $\bar{h}_{\mu\nu} = h_{\mu\nu} - \frac{1}{2} \eta_{\mu\nu} h$ is the trace-reversed metric perturbation. For the time-time component, this reduces to a Poisson-like equation sourcing Newtonian potential perturbations. In this context, localized regions of negative energy produce curvature fluctuations characterized by localized reductions in scalar curvature \(R\), Ricci tensor eigenvalues, and geodesic deviation.

The combined influence of QET array parameters and experimental enhancements on the signal-to-noise ratio (SNR) can be effectively modeled by the approximate relation

\begin{equation}
\mathrm{SNR} \sim \left( \frac{N}{d^3} \cdot \frac{F}{\pi} \cdot G_{\mathrm{ent}} \cdot G_{\mathrm{shape}} \cdot G_{\mathrm{multi}} \right) \Bigg/ \left( \frac{1}{\sqrt{f}} \cdot e^{-r} \cdot G_{\mathrm{noise}} \right),
\label{eq:snr_model}
\end{equation}
\DSZ 

where \(N\) is the number of QET units, \(d\) the detector spacing, \(F\) the cavity finesse, and \(f\) the repetition rate. The dimensionless multipliers \(G_{\mathrm{ent}}, G_{\mathrm{shape}}, G_{\mathrm{multi}}\) represent entanglement depth, field shaping, and multimode coupling enhancements, while \(r\) and \(G_{\mathrm{noise}}\) quantify squeezing and environmental noise suppression.

Full derivation of Eq.~\ref{eq:snr_model}, including Gaussian modeling of the curvature response and Green’s function integration of the semiclassical Einstein equations, is provided in Appendix~\ref{appendix:snr_model}.

\begin{figure}[htbp]
\centering
\includegraphics[width=\linewidth]{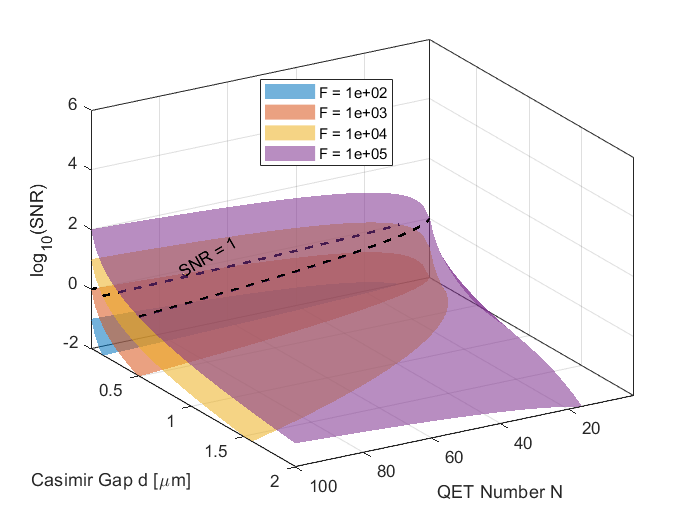} 
\vspace{-0.2cm}
\caption{
Contour plot of \(\log_{10}(\mathrm{SNR})\) as a function of the number of QET units \(N\) and detector spacing \(d\). 
Different curves or surface colors represent varying cavity finesse \(F\) values, demonstrating how increased finesse significantly enhances the signal-to-noise ratio.
The shaded region below the dashed contour line (SNR = 1) indicates parameter regimes where detection is not feasible with current technology.
This visualization guides optimization of detector array parameters to maximize curvature signal detectability.
}
\label{fig:Fig2_snr_contour}
\vspace{-0.2cm}
\end{figure}

Figure \ref{fig:Fig2_snr_contour} visualizes \(\log_{10}(\mathrm{SNR})\) computed from Eq.~\ref{eq:snr_model} over sweep variables \(N\) and \(d\), highlighting parameter regimes conducive to detection for various cavity finesse values \(F\). This framework guides optimization by revealing how different mechanisms collectively boost the curvature signal above noise floors.

While this model captures the dominant parametric dependencies, a full experimental noise budget includes technical noise sources and decoherence effects not explicitly shown here. 
which can be addressed in a detailed engineering study.

The resulting curvature signal $\delta R(x,t)$ depends not only on the energy density amplitude but also on its localization, duration, and detector response function. 

Table~\ref{tab:EnhancementNoise} outlines qualitative mechanisms for boosting curvature signals and their associated vulnerabilities. Table~\ref{tab:SNRModel} then models the core parameters that govern these effects quantitatively across different experimental regimes.

\begin{table}
\caption{Qualitative signal enhancement strategies in QET-based negative energy experiments, including the dominant noise sources and net impact on signal-to-noise ratio (SNR).}
\label{tab:EnhancementNoise}
\resizebox{\textwidth}{!}{%
\begin{tabular}{llll}
\br
\textbf{Mechanism} & \textbf{Signal Enhancement Effect} & \textbf{Dominant Noise Sources} & \textbf{Impact on SNR} \\
\mr
Gate synchronization & Temporal narrowing of $\langle T_{00} \rangle$ & Clock jitter, phase drift & Improves localization; sensitive to $\Delta t$ \\
Detector proximity ($d \to 0$) & Boosts field overlap & Cross-talk, backreaction & Enhances constructive interference \\
Spatial smearing design & Optimizes field–detector coupling & Beam shaping imperfections & Improves robustness to misalignment \\
Postselection filtering & Amplifies per-event signal & Reduced event rate, sampling bias & Raises per-trial SNR; reduces throughput \\
Chained QET steps & Accumulates curvature from multiple Bobs & Thermal noise, decoherence & Enables additive signal amplification \\
\br
\end{tabular}
}
\end{table}
\vspace{0.2cm}
\DSZ 

\begin{table}
\caption{Parametric model for optimizing the curvature signal-to-noise ratio (SNR) in QET-based setups. This symbolic summary identifies the primary factors affecting detectability. The main sweep parameters are detector number $N$ and spacing $d$, while additional toggles influence signal strength or noise floor. For specific simulation values used in numerical plots, see Table~\ref{tab:Fig2_params}.}
\label{tab:SNRModel}
\resizebox{\textwidth}{!}{%
\begin{tabular}{lll}
\br
\textbf{Parameter / Toggle} & \textbf{Channel Affected} & \textbf{SNR or $\delta R$ Dependence} \\
\mr
$N$ (number of detectors) & Signal accumulation & SNR $\propto N$ (linear); coherent sum boosts to $N^2$ in ideal case \\
$d$ (detector separation) & Signal localization vs overlap & Optimal at $d \sim \sigma$; large $d$ reduces field coherence \\
$r$ (squeezing parameter) & Signal amplitude & SNR $\propto e^{2r}$; boosts $\langle T_{00} \rangle$ and $\delta R$ \\
$\Delta t$ (timing jitter) & Noise floor & SNR $\propto 1/\Delta t$; affects effective pulse duration \\
$\sigma$ (spatial width) & Signal shape & Controls localization and coupling bandwidth \\
$\gamma$ (decoherence rate) & Noise floor & Reduces entanglement transfer efficiency \\
\br
\end{tabular}
}
\end{table}
\DSZ 
\begin{table}[ht]
\centering
\caption{Simulation parameters used to generate the finesse-sweep SNR surface plot shown in Figure~\ref{fig:Fig2_snr_contour}. SNR was computed across a parameter grid of QET number $N$ and Casimir gap $d$, incorporating enhancement factors and realistic noise models.}
\label{tab:Fig2_params}
\begin{tabular}{|l|c||l|c|}
\hline
\textbf{Parameter} & \textbf{Value} & \textbf{Parameter} & \textbf{Value} \\
\hline
Squeezing parameter $r$ & 1.5 & QET repetition rate $f$ & $10^5$ Hz \\
Cavity finesse values $F$ & $10^2$–$10^5$ & Temperature $T$ & 300 K \\
Entanglement gain $G_{\mathrm{ent}}$ & 2 & Mass $m$ & $1\,\mu$g \\
Shaping factor $G_{\mathrm{shape}}$ & 2 & Quality factor $Q$ & $10^6$ \\
Multimode gain $G_{\mathrm{multi}}$ & 2 & Technical noise floor & $10^{-3}$ \\
Detector noise floor & $10^{-4}$ & Gap noise model & $10^{-2}/(d\, [\mu\mathrm{m}])$ \\
\hline
\end{tabular}
\end{table}


The amplitude of the induced curvature scales with the energy density $\epsilon$ and inversely with the square of the localization width. Enhanced localization via detector synchronization and squeezing of the initial field state can significantly amplify the signal. In subsequent sections, we explore how these parameters can be tuned experimentally to optimize sensitivity.

\section{Observable Signatures and Sensitivity} \label{sec:ObsSigSen_5}

To establish the experimental viability of detecting QET-induced negative energy, we evaluate observable consequences of its semiclassical gravitational backreaction. We focus on three independent detection modalities: interferometric phase shifts, inertial responses, and relativistic clock drift. Each responds to spacetime curvature \(\delta R\) with distinct spatial and temporal sensitivities.

\subsection{Target Observables}

The primary observables accessible to current and near-future instruments are:

\paragraph*{Optical Path Shifts}  
A localized curvature dip alters the spacetime interval along interferometer arms, producing a phase shift \(\Delta \phi\) proportional to the curvature perturbation \(\delta R\) and the interferometer arm length \(L\). For weak fields and small curvature, the path length change \(\Delta L\) can be approximated as \(\Delta L \sim \delta R \cdot L^2\), yielding

\begin{equation}
    \Delta \phi = \frac{2\pi}{\lambda} \Delta L \approx \frac{2\pi}{\lambda} \delta R \, L^2,
\end{equation}
\DSZ 

where \(\lambda\) is the optical wavelength.  
\DSZ 

\paragraph*{Inertial Acceleration}  
MEMS accelerometers placed near the QET interaction site register deviations \(\Delta a\) in proper acceleration arising from local curvature gradients. These inertial effects may be enhanced by resonance or synchronized gating, providing a complementary detection channel \cite{Kolkowitz2016}.

\paragraph*{Atomic Clock Drift}  
Transient curvature pulses modify the gravitational potential \(\Phi\), inducing proper time shifts \(\Delta \tau / \tau\) detectable by optical lattice clocks with fractional frequency stability below \(10^{-18}\) \cite{ludlow2015optical}. The clock frequency shift relates to the potential difference via

\begin{equation}
    \frac{\Delta f}{f} \sim \frac{\Delta \Phi}{c^2} \sim \frac{\delta R \, L^2}{c^2},
\end{equation}
\DSZ 

where \(c\) is the speed of light.

\subsection{Sensitivity Scaling with Entanglement and Integration Time}

In the semiclassical weak-field regime, the curvature perturbation induced by a collection of \(N\) entangled QET units scales approximately linearly:
\begin{equation}
    \delta R(N) \sim N \cdot \delta R_0,
\end{equation}
\DSZ 
where \(\delta R_0 \sim 10^{-36}~\mathrm{m}^{-2}\) corresponds to the curvature from a single QET pair, estimated from typical energy densities generated in the protocol.

Repeated measurements and integration over a total duration \(T\) reduce statistical noise assuming uncorrelated fluctuations. The effective detection threshold improves as \(1/\sqrt{T}\), enabling sensitivity to sub-threshold curvature with sufficient averaging.

\begin{figure}[H]
\centering
\includegraphics[width=0.95\linewidth]{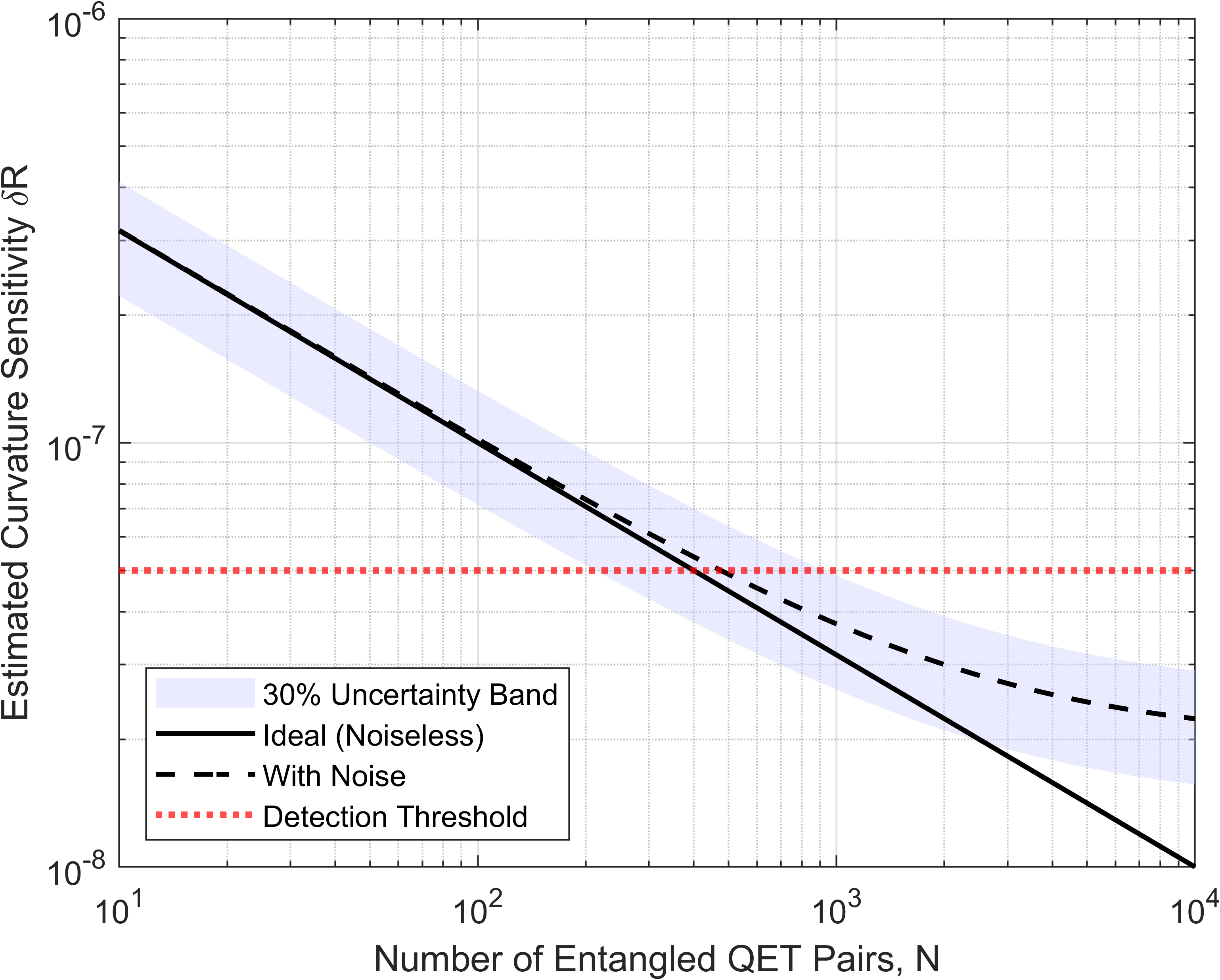}
\caption{
Estimated curvature sensitivity \(\delta R\) versus the number of entangled QET pairs \(N\). The ideal scaling (solid) assumes noiseless additive accumulation. The dashed curve includes thermal and optical noise floors. The shaded region represents a 30\% uncertainty band, and the red dotted line denotes a representative detection threshold.
}
\label{fig:Fig3_deltaR_xt_qet_array}
\end{figure}
\DSZ 

\subsection{Signal-to-Noise Ratio and Detection Thresholds}

The signal-to-noise ratio (SNR) for curvature-based detection is defined as
\begin{equation}
    \mathrm{SNR} = \frac{\delta R(N)}{\sigma_R},
\end{equation}
where \(\sigma_R\) is the root-mean-square curvature noise intrinsic to the detector.

For various platforms, the noise term \(\sigma_R\) takes distinct functional forms:

\paragraph*{Interferometry}
Shot-noise-limited interferometers yield curvature noise approximately

\begin{equation}
    \sigma_R \sim \frac{\lambda}{L^2} \cdot \frac{1}{\sqrt{T}} \cdot \frac{1}{\mathrm{SNR}_\mathrm{opt}},
\end{equation}
\DSZ 

where \(L\) is the interferometer arm length, \(\lambda\) is the laser wavelength, and \(\mathrm{SNR}_\mathrm{opt}\) characterizes optical detection sensitivity. Typical values are \(L \sim 1~\mathrm{m}\) and \(\lambda \sim 10^{-6}~\mathrm{m}\).

\paragraph*{MEMS-Based Inertial Sensors}
The minimal detectable curvature noise is related to acceleration sensitivity \(\delta a_{\min}\) and sensor scale \(L\) by
\begin{equation}
    \sigma_R \sim \frac{\delta a_{\min}}{L},
\end{equation}
with \(\delta a_{\min} \sim 10^{-10}~\mathrm{m/s^2/\sqrt{Hz}}\) and \(L \sim 1~\mathrm{mm}\) \cite{Kolkowitz2016}.

\paragraph*{Clock-Based Detection}
Atomic clock frequency shifts relate directly to curvature by
\begin{equation}
    \frac{\Delta f}{f} \sim \frac{\Phi}{c^2} \sim \frac{\delta R \, L^2}{c^2}.
\end{equation}

\DSZ 

\subsection{Scaling of SNR with Number of QET Units}

\begin{figure}[htbp]
\centering
\includegraphics[width=0.95\linewidth]{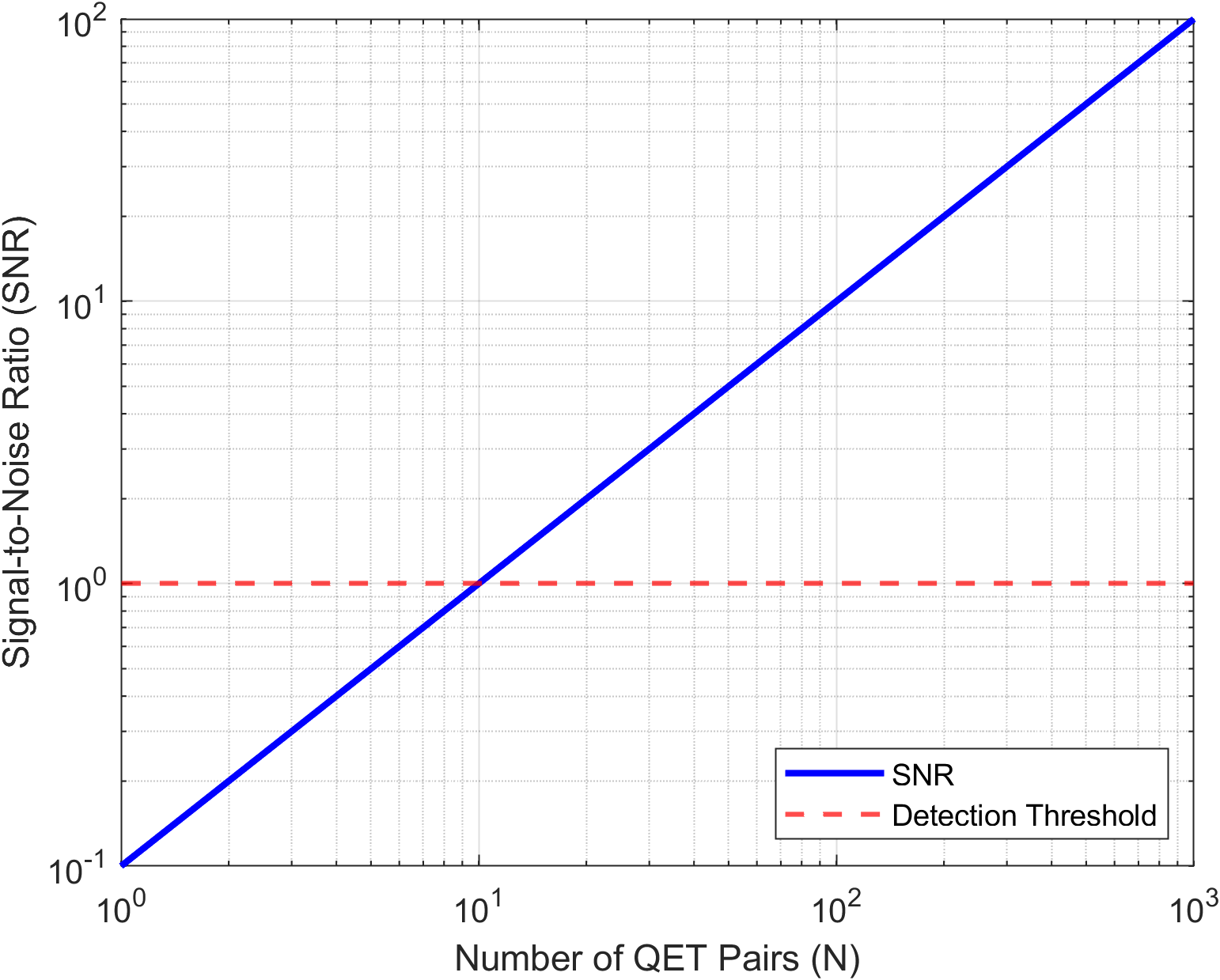}
\caption{Log-log scaling of signal-to-noise ratio (SNR) as a function of entangled QET pairs \(N\). Detection threshold at \(\mathrm{SNR} = 1\) shown in red dashed line.}
\label{fig:Fig6_SNRvsN}
\end{figure}

Figure~\ref{fig:Fig6_SNRvsN} shows the log-log scaling of the SNR with the number of entangled QET units \(N\), assuming a representative noise floor \(\sigma_R \sim 10^{-35}~\mathrm{m}^{-2}\). Detection requires \(\mathrm{SNR} > 1\), achievable with modest arrays of \(N \sim 10 - 100\) entangled units using current or near-future technology. These SNR estimates are derived assuming curvature amplitudes consistent with the spatial profiles in Figure~\ref{fig:Fig5_clock_drift_vs_deltaR}.

\DSZ 

\subsection{Spatial Curvature Profiles from QET Architectures}

The spatial distribution of negative energy directly shapes the induced curvature profile via the linearized Einstein equation in the weak static regime:
\begin{equation}
    \delta R(x) = -8\pi G \, \langle T_{00}(x) \rangle,
    \label{eq:R_from_T00}
\end{equation}
where \(G\) is Newton's gravitational constant.

\begin{figure}[htbp]
    \centering
     \includegraphics[width=0.95\linewidth]{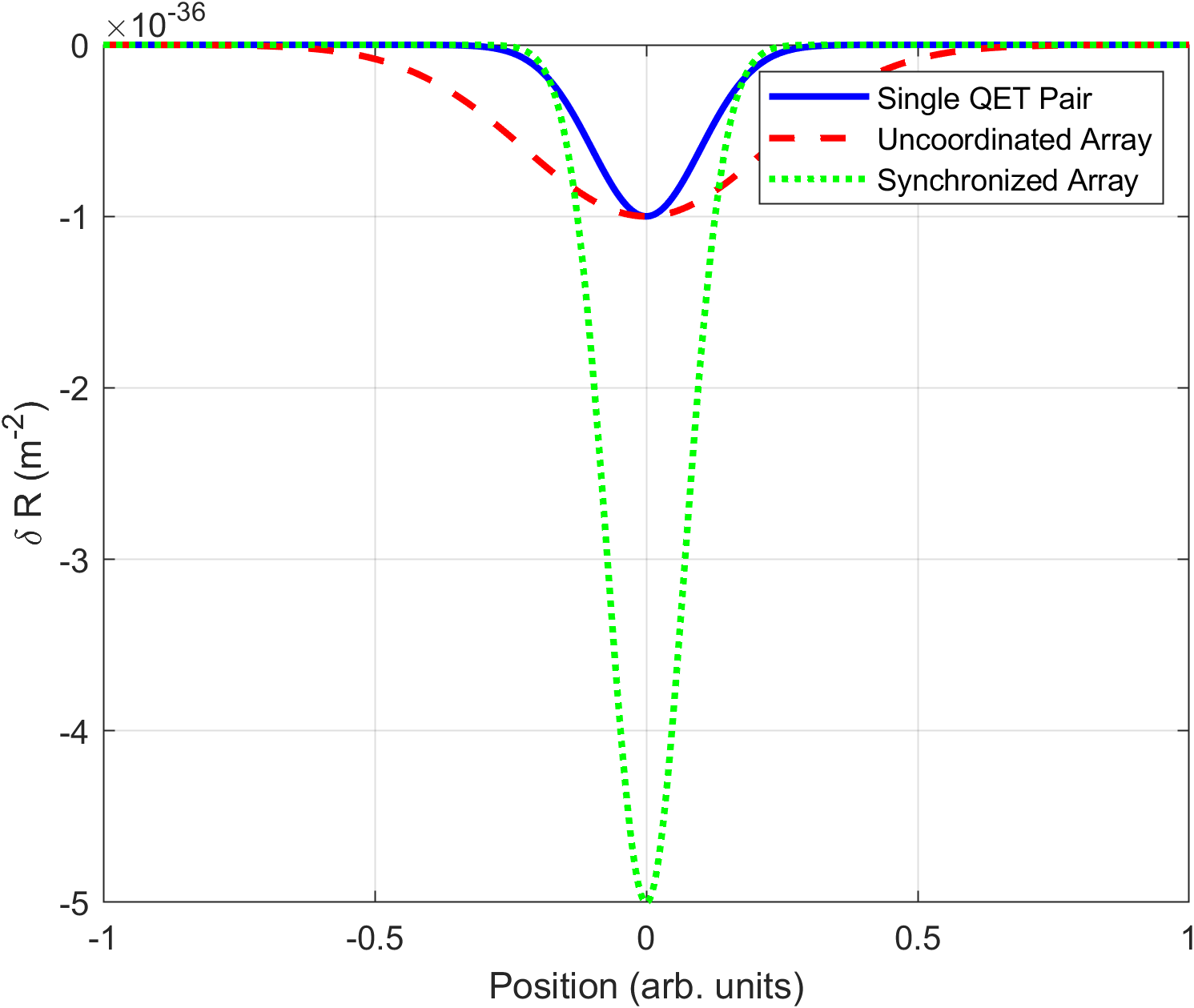}
    \caption{Simulated curvature profiles \( \delta R(x) \) generated by different QET architectures: single pair, unsynchronized array, and synchronized array. Derived from localized Gaussian models of \( \langle T_{00}(x) \rangle \). See Appendix~\ref{appendix:curvature_profiles} for modeling details.}
    \label{fig:Fig5_clock_drift_vs_deltaR}
\end{figure}

Architectural configurations of QET units determine both the amplitude and localization of \(\delta R(x)\). Figure~\ref{fig:Fig5_clock_drift_vs_deltaR} illustrates representative profiles for single pairs, unsynchronized arrays, and synchronized arrays, modeled as localized Gaussians with parameters derived from detector geometry and coupling strengths. These profiles serve as inputs to the SNR and detection analyses above.

\begin{table*}[htbp]
\caption{Simulated curvature profiles and corresponding signal-to-noise ratio (SNR) estimates from representative QET configurations. Peak curvature \(\delta R_{\mathrm{peak}}\) and spatial width \(\Delta x\) influence detectability.}
\label{tab:qet_curvature_profiles}
\centering
\renewcommand{\arraystretch}{1.3}
\begin{tabular}{lccc}
\br
Architecture & \(\delta R_{\mathrm{peak}}\,(\mathrm{m}^{-2})\) & Width \(\Delta x\) (m) & Estimated SNR \\
\mr
Single QET Pair & \(-1 \times 10^{-36}\) & 0.2 & Low \\
Uncoordinated Array & \(-1 \times 10^{-36}\) & 0.4 & Moderate \\
Synchronized Array & \(-5 \times 10^{-36}\) & 0.1 & High \\
\br
\end{tabular}
\end{table*}
\DSZ 

\section{Concentrating Negative Energy via Superposition}\label{sec:ConNegEnSup_6}

To maximize the semiclassical gravitational imprint of quantum energy teleportation (QET), it is not sufficient to merely generate negative energy density—one must also shape it in space and time. This section presents architectures that engineer the stress-energy profile using interference-based spatial localization, time-domain pulse shaping, and detector-specific enhancement strategies.

\subsection{Interference-Based Spatial Concentration}

When a quantum system is placed in a spatial superposition—such as a matter-wave interferometer—each branch of the wavefunction can be independently coupled to a localized QET operation. Upon coherent recombination, the resulting interference reshapes the energy density \(\langle T_{00}(x) \rangle\), concentrating negative energy at specific spatial locations.

Consider a system prepared in the symmetric superposition:

\[
|\psi\rangle = \frac{1}{\sqrt{2}}\left(|L\rangle + |R\rangle\right),
\]
\DSZ 

where each branch \(L\) and \(R\) interacts with the vacuum field via a local QET protocol. The resulting energy density profile is approximately:
\begin{equation}
\langle T_{00}(x) \rangle \sim -\lambda \left[ \delta(x - x_L) + \delta(x - x_R) \right] + I(x),
\end{equation}
where \(\lambda > 0\) characterizes the energy scale of each QET interaction, and \(I(x)\) contains interference cross-terms dependent on the relative phase and spatial overlap of the branches. Constructive interference can thus localize negative energy at the recombination point.

\begin{figure}[H]
\centering
\includegraphics[width=0.95\linewidth]{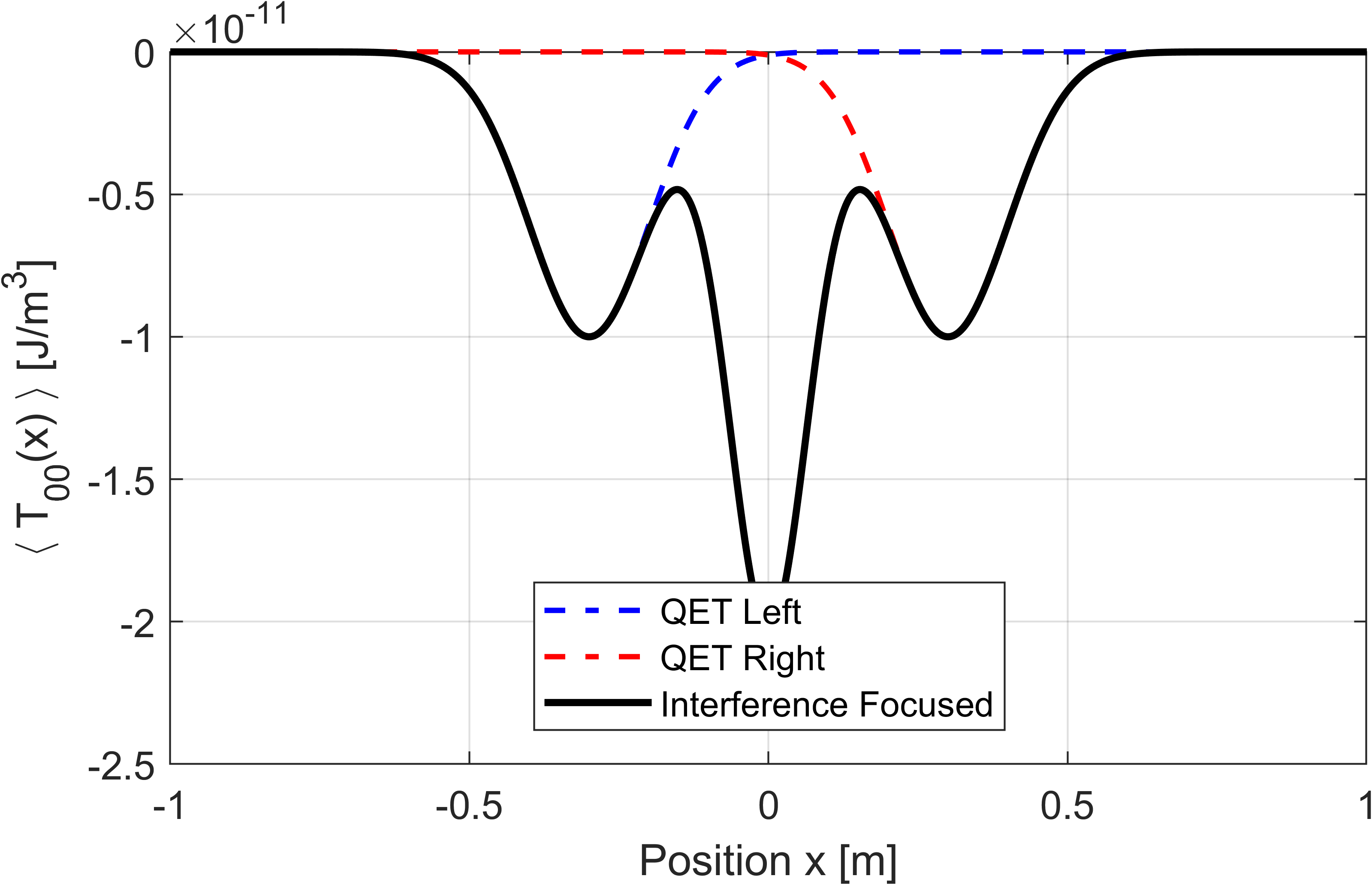}
\caption{Quantum superposition of two QET channels. Branches \(L\) and \(R\) interfere coherently after independent QET operations, focusing negative energy density at the recombination point.}
\label{fig:Fig6_interference_qet}
\end{figure}
\DSZ 

Such interference-based control enables engineered curvature sources through entangled field manipulation, tunable via phase, geometry, and pulse timing.

\subsection{Time-Gated Curvature Pulses}

Unlike static Casimir setups, QET operations can be activated on demand. Applying QET operations in synchrony across an array produces a time-localized curvature dip, effectively generating a gravitational pulse detectable via optical phase shifts or clock drift.

We model the scalar curvature pulse as a Gaussian function of time:

\begin{equation}
\delta R(t) = N \cdot \delta R_0 \cdot \exp\left[ -\frac{(t - t_0)^2}{2\sigma^2} \right],
\end{equation}
\DSZ 

where \(N\) is the number of synchronized QET pairs, \(\delta R_0\) is the curvature contribution per pair, and \(\sigma\) controls the pulse temporal width.

The induced optical path length shift \(\Delta L(t)\) for an interferometer with baseline \(L_0\) is approximated as

\begin{equation}
\Delta L(t) \approx \frac{1}{2} \delta R(t) \cdot L_0 \cdot t^2,
\end{equation}
\DSZ 

valid in the weak-field, linearized curvature regime.

\begin{figure*}[t]
\centering
\begin{subfigure}[t]{0.48\textwidth}
\centering
\includegraphics[width=\linewidth]{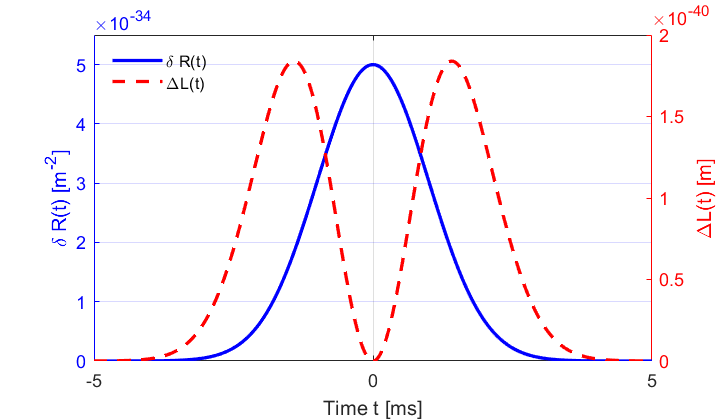}
\caption{Time-resolved curvature pulse \(\delta R(t)\) (blue) and induced interferometer path shift \(\Delta L(t)\) (red).}
\end{subfigure}
\hfill
\begin{subfigure}[t]{0.48\textwidth}
\centering
\includegraphics[width=\linewidth]{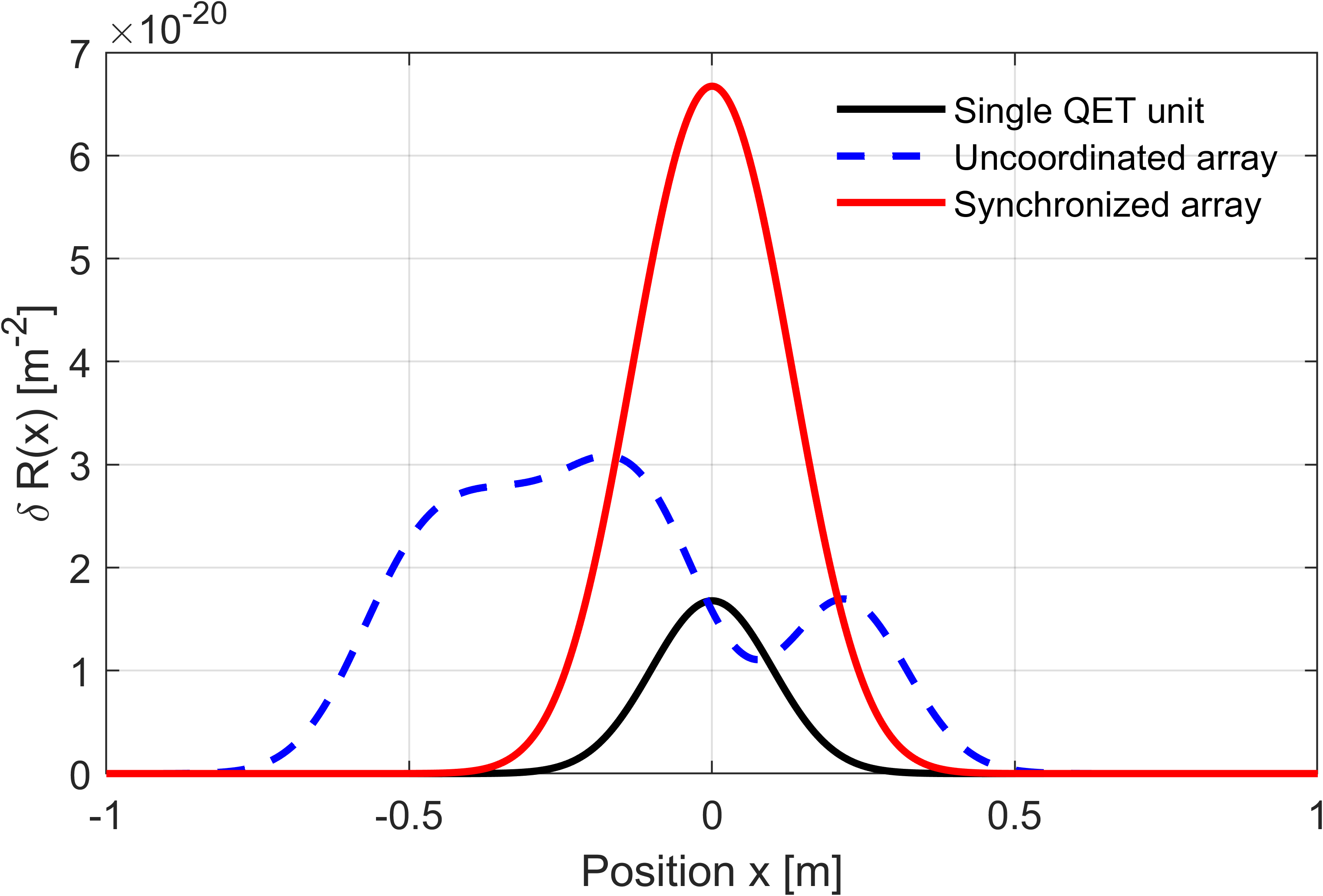} 
\DSZ 
\caption{Spatial curvature profiles for QET arrays. Synchronization enhances peak depth and spatial localization.}
\end{subfigure}
\caption{Temporal and spatial engineering of negative energy. QET pulses can be focused in time and space for enhanced gravitational detectability.}
\label{fig:Fig7_curvature_pulse_profiles}
\end{figure*}

\subsection{Atomic Clock Drift Detection}

Localized curvature induces proper time dilation measurable by high-precision atomic clocks. The fractional clock time shift during a curvature pulse of duration \(\Delta t\) is approximately
\begin{equation}
\frac{\Delta \tau}{\tau} \approx \frac{1}{12} \delta R \cdot L^2 \cdot \Delta t,
\end{equation}
where \(L\) is the spatial extent over which curvature is significant, and \(\delta R\) its magnitude.

Current state-of-the-art optical clocks reach fractional frequency stability near \(10^{-18}\), making QET-induced pulses potentially detectable if the number of synchronized QET units \(N\) is sufficiently large.

Table~\ref{tab:SNRparams} lists the parameter values used to generate the SNR surface in Figure~\ref{fig:Fig11_snr_contour}. The detection threshold contour at $\log_{10}(\mathrm{SNR}) = 0$ corresponds to a baseline configuration of $N_0 = 10^3$ entangled pairs with separation $d_0 = 0.05$ m.

\begin{table}[ht]
\centering
\caption{Parameter values used in the log-scaled SNR contour plot (Figure~\ref{fig:Fig11_snr_contour}). The detection threshold is defined at \(N_0 = 10^3\), \(d_0 = 0.05\,\mathrm{m}\).}
\label{tab:SNRparams}
\begin{tabular}{|l|c||l|c|}
\hline
\textbf{Parameter} & \textbf{Value} & \textbf{Parameter} & \textbf{Value} \\
\hline
Squeezing parameter $r$ & 1.5 & Entanglement gain $G_{\mathrm{ent}}$ & 10 \\
Cavity finesse $F$ & $10^4$ & Spatial shaping factor $G_{\mathrm{shape}}$ & 5 \\
QET repetition rate $f$ & $10^5$ Hz & Multimode enhancement $G_{\mathrm{multi}}$ & 3 \\
Reference QET count $N_0$ & $10^3$ & Noise shielding $G_{\mathrm{noise}}$ & 1 \\
Reference spacing $d_0$ & 0.05 m & — & — \\
\hline
\end{tabular}
\end{table}

\begin{figure}[H]
\centering
\includegraphics[width=0.95\linewidth]{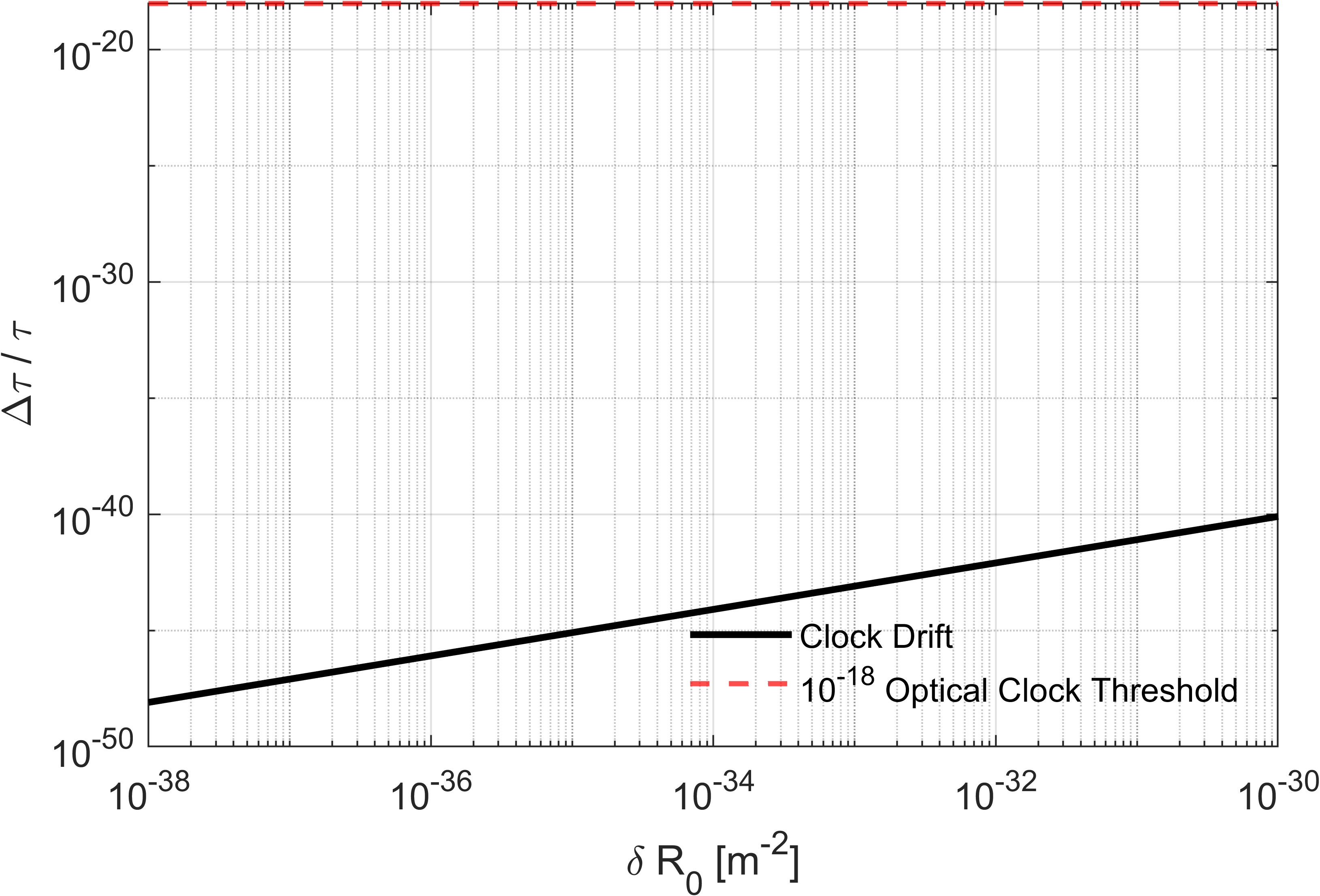}
\caption{Fractional clock drift \(\Delta \tau / \tau\) as a function of curvature amplitude \(\delta R_0\) for fixed spatial extent \(L = 1\,\mathrm{mm}\) and pulse duration \(\Delta t = 1\,\mathrm{ms}\). The curves indicate the approximate detectability thresholds for optical atomic clocks.}
\label{fig:Fig8_clock_drift_vs_deltaR}
\end{figure}
\DSZ 
\subsection{Strain Detection Analogy}

The curvature pulses generated by QET induce differential strain \(h(t)\) across interferometer arms, analogous to gravitational waves, estimated as
\begin{equation}
h(t) \sim \frac{1}{2} \delta R(t) \cdot L^2,
\end{equation}
where \(L\) is the arm length.

While QET-induced strain amplitudes \(h_{\mathrm{QET}} \sim 10^{-35}\) are many orders of magnitude below current LIGO sensitivity, integrating over multiple pulses or employing cryogenic interferometer scaling may improve detectability.

\begin{figure}[H]
\centering
\includegraphics[width=0.95\linewidth]{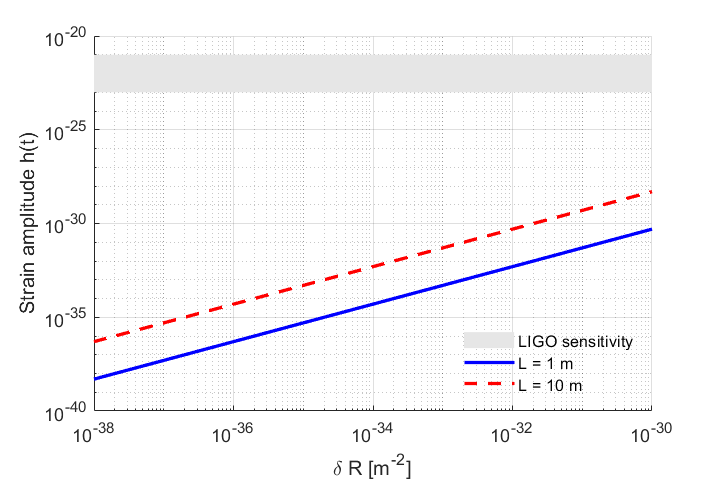}
\caption{Strain amplitude \(h(t)\) versus curvature amplitude \(\delta R\) for interferometer baselines \(L=1\,\mathrm{m}\) and \(L=10\,\mathrm{m}\). The shaded region denotes the current LIGO sensitivity band for gravitational-wave detection.}
\label{fig:Fig9_strain_vs_curvature}
\end{figure}

\subsection{Consolidated Clock Shift Results}

\begin{figure}[H]
\centering
\includegraphics[width=0.95\linewidth]{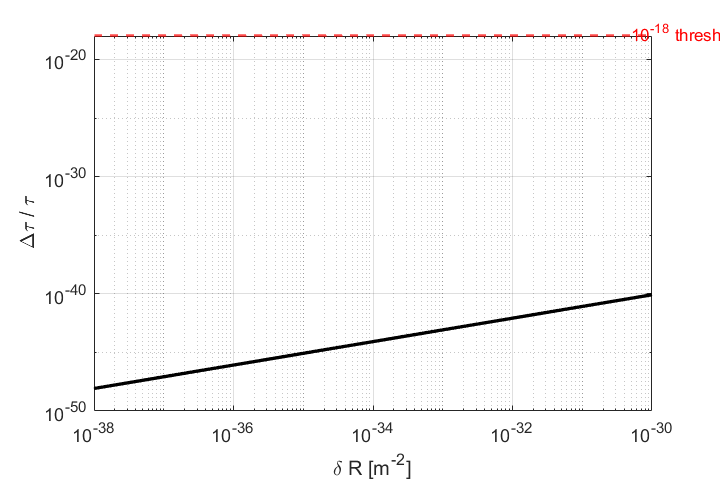}
\caption{Simulated fractional clock drift \(\Delta \tau / \tau\) versus curvature pulse amplitude \(\delta R\). Parameters: spatial extent \(L=1\,\mathrm{mm}\), pulse duration \(\Delta t=1\,\mathrm{ms}\).}
\label{fig:Fig10_clock_drift_combined}
\end{figure}

\begin{figure}[H]
\centering
\includegraphics[width=0.95\linewidth]{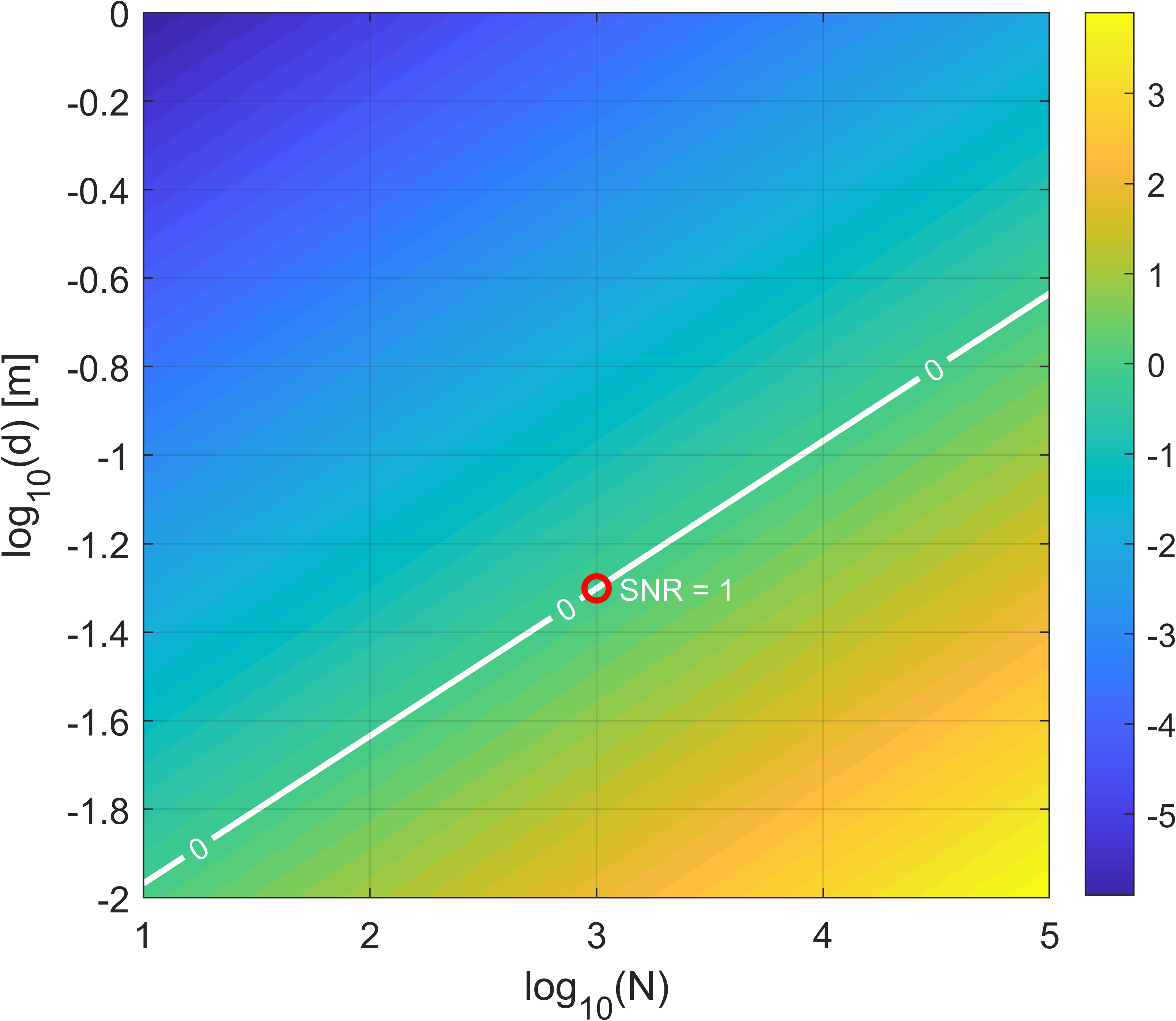}
\caption{Simulated log-log SNR as a function of QET unit number \(N\) and spacing \(d\), assuming fixed squeezing \(r=1.5\), cavity finesse \(F=10^4\), and repetition rate \(f = 10^5\) Hz. Detection threshold at \(\log_{10}(\mathrm{SNR}) = 0\) is shown as a white contour.}
\label{fig:Fig11_snr_contour}
\end{figure}

Combining theoretical and numerical results, Figure~\ref{fig:Fig10_clock_drift_combined} illustrates the fractional clock drift \(\Delta \tau / \tau\) as a function of curvature pulse \(\delta R\), assuming fixed \(L=1\,\mathrm{mm}\) and \(\Delta t=1\,\mathrm{ms}\). This highlights practical detection thresholds and the scaling behavior relevant for experimental design.

\subsection{Summary}

QET enables not only the generation of negative energy but also its targeted concentration. Through quantum interference, pulse timing, and detector-specific coupling, spacetime perturbations can be engineered for maximal gravitational imprint. These strategies are essential for surpassing current sensitivity limits and realizing laboratory-scale curvature detection.

\section{Sensitivity and Feasibility} \label{sec:SensFeas_7}

To assess the practical reach of QET-induced curvature detection, we analyze both theoretical sensitivity scaling and experimental feasibility across multiple detection strategies. This section integrates numerical simulations, detection thresholds, and realistic parameter sweeps to chart the achievable signal-to-noise ratio (SNR) under experimentally relevant conditions.

\subsection*{A. Parameter Scaling and Sensitivity Contours}

Figure~\ref{fig:Fig2_snr_contour} illustrates the logarithmic SNR landscape as a function of two dominant parameters: the number of QET units \(N\) and their spacing \(d\). The simulated signal incorporates enhancements from cavity finesse \(F\), entanglement depth, spatial shaping, and multimode structure, while the noise includes effects from squeezing, repetition, and shielding.

As shown, increasing the array size \(N\) and reducing inter-detector spacing \(d\) yields a superlinear gain in SNR, consistent with the scaling:
\[
\mathrm{SNR} \sim \left( \frac{N}{d^3} \cdot \frac{F}{\pi} \cdot G_{\mathrm{ent}} \cdot G_{\mathrm{shape}} \cdot G_{\mathrm{multi}} \right) \Bigg/ \left( \frac{1}{\sqrt{f}} \cdot e^{-r} \cdot G_{\mathrm{noise}} \right).
\]

Contour levels crossing \(\log_{10}(\mathrm{SNR}) \sim 0\) identify the detection threshold. These results suggest that laboratory-scale QET arrays with modest finesse and cryogenic shielding can surpass detectability thresholds within current or near-term technologies.

\DSZ 

\subsection*{B. Comparative Detection Strategies}

We now compare the sensitivity of interferometric, clock-based, and strain-based detection approaches under optimized QET configurations. Table~\ref{tab:comparison} summarizes the characteristic observables, expected signals, noise limits, and scalability.

\begin{table*}[t]
\centering
\resizebox{\linewidth}{!}{
\begin{tabular}{|l|c|c|c|c|c|c|}
\hline
\textbf{Detection Mode} & \textbf{Observable} & \textbf{Expected Signal} & \textbf{Sensitivity Threshold} & \textbf{Enhancing Factors} & \textbf{Noise Sources} & \textbf{Feasibility} \\
\hline
Atomic Clocks & \(\Delta \tau/\tau\) & \(10^{-21} \mathrm{~to~} 10^{-17}\) & \(\sim 10^{-18}\) & Co-location, repetition, squeezing & Drift, calibration & High (lab-scale) \\
Interferometry & \(\Delta L(t)\) & 
\(10^{-20} \mbox{ to } 10^{-15}~\mbox{m}\) & \(10^{-18}~\mathrm{m}\) & Finesse, multimode shaping & Shot noise, thermal & Medium (optical cavity) \\
Strain (GW-like) & \(h(t)\) & 
\(10^{-35}~\mathrm{to}~10^{-30}\)
 & \(10^{-23}\) (LIGO) & Array coherence, cryogenics & Quantum backaction & Low (next-gen) \\
\hline
\end{tabular}
}
\caption{Comparison of detection strategies for QET-induced curvature. Clock drift is the most feasible method at present precision levels.}
\label{tab:comparison}
\end{table*}
\DSZ 

While interferometry benefits from high optical finesse, it is limited by noise floors unless enhanced via quantum squeezing and repetition. Strain-based detection lags due to its extremely small amplitude but may become viable in next-generation cryogenic setups. Clock-based detection, leveraging atomic stability at \(10^{-18}\) and beyond, currently offers the most accessible sensitivity.

\subsection*{C. Scalability and Noise Mitigation}

Scalability is dominated by three key parameters: energy localization, coherence maintenance, and environmental suppression. The QET protocol allows deterministic timing and shaping of energy profiles, unlike Casimir-based static systems. Moreover, quantum squeezing and repetition can further suppress technical noise without amplifying curvature backreaction.

\DSZ 

Near-term implementations might deploy optomechanical arrays or entangled ion traps in vacuum chambers, with spacing \(d \sim 0.1\,\mathrm{m}\), squeezing \(r \gtrsim 1.5\), and finesse \(F \sim 10^4\). Numerical results suggest such configurations can cross the SNR threshold with \(\mathcal{O}(10^4)\) shots or integration over seconds.

\subsection*{D. Summary of Sensitivity Landscape}

Figure~\ref{fig:Fig10_clock_drift_combined} and Table~\ref{tab:comparison} jointly illustrate that QET-induced curvature may be observable via atomic clock drift using current technology. Interferometry provides complementary access with optical shaping, while strain detection remains future-facing. The parametric contours in Figure~\ref{fig:Fig2_snr_contour} reveal accessible regions of experimental phase space, guiding both design and scaling.


\section{Beyond Warp Bubbles: Quantum–Curvature Compression}
\label{sec:QIXC_8}

Building on the signal enhancement strategies and curvature localization mechanisms developed in Sec.~\ref{sec:FocNegEnergy_4}, we now explore a speculative extension: the possibility of coherently steering and propagating localized curvature pulses. Rather than solely amplifying the detectability of negative energy, we consider whether QET-based architectures can transiently shape spacetime curvature in a directed manner. These structures, which we refer to as quantum–curvature compression or QIX specialized for curvature compression (QIX-C), aim to produce traveling zones of reduced scalar curvature, analogous in some respects to the engineered geometries proposed in warp drive scenarios \cite{Alcubierre1994} but rooted in causal, field-theoretic dynamics.

\subsection{Overview and Motivation}
\label{subsec:QIXC_overview}

Proposals such as the Alcubierre warp metric or Natário’s generalized constructs involve superluminal effective transport by embedding a flat region inside a shell of exotic stress–energy violating the null energy condition. While these geometries are solutions to Einstein’s equations, their reliance on static, fine-tuned negative energy distributions makes them physically implausible in semiclassical settings. 

By contrast, the QIX-C framework does not require static configurations or global topology change. Instead, it utilizes dynamically generated negative energy pulses from synchronized QET arrays to induce transient, traveling curvature dips. These propagating compression channels do not enable superluminal travel but may provide a controllable way to modulate proper time intervals, curvature gradients, or geodesic deviation in a confined region. Such constructs could potentially serve as testbeds for gravitationally mediated quantum signaling, spacetime squeezing, or lensing effects.

\subsection{QIX-C Geometry and Curvature Compression}
\label{subsec:QIX-C_geometry}

We define a QIX-C as a spacetime region where a negative energy pulse, generated via a temporally and spatially gated QET protocol, induces a transient reduction in Ricci scalar curvature:
\begin{equation}
    \delta R(x,t) \approx -\epsilon\, \exp\left( -\frac{(x - vt)^2}{2\sigma^2} \right),
    \label{eq:QIX-C_profile}
\end{equation}
where \(\epsilon > 0\) sets the magnitude of the curvature dip, \(v\) is the effective propagation speed of the pulse (sub-luminal, controlled by sequential QET timing), and \(\sigma\) governs the spatial localization.

This structure can be extended to 3+1 dimensions by replacing \(x \to \mathbf{x}\cdot \hat{n}\) and introducing transverse falloff. The dynamics of the curvature pulse follow from the semiclassical Einstein equations:
\begin{equation}
    \Box \delta h_{\mu\nu}(x) = -16\pi G \, 
     \langle T_{\mu\nu}(x) \rangle_{\mathrm{QET}},
\end{equation}
\DSZ 

with 
\(\langle T_{\mu\nu} \rangle_{\mathrm{QET}}\)
 constructed from a sequence of temporally staggered energy injections with net negative integrated energy, respecting quantum inequalities.

\subsection{Simulation Results and Compression Dynamics}
\label{subsec:QIX-C_simulations}

To explore feasibility, we simulate QIX-C formation using a chain of \(N\) QET interactions spaced by distance \(d\), with synchronized gate functions producing overlapping negative energy lobes. Each event is modeled using a Gaussian \(\langle T_{00}(x,t) \rangle\), and the resulting curvature profile \(\delta R(x,t)\) is obtained using retarded Green's functions in 1+1 dimensions.

Results (Fig. \ref{fig:Fig13qixprofile}) show that when the timing offsets and spatial separations are tuned such that individual curvature dips overlap constructively, a traveling wave packet of negative curvature forms. The pulse maintains spatial localization and moves at a controlled group velocity ($v_{\mathrm{eff}}$), determined by the timing of the QET gates. The induced compression is strongest when the energy profiles interfere coherently and the repetition rate matches the propagation velocity.

\subsection{Figures and Visualizations}
\label{subsec:QIX-C_figures}


\begin{figure}[ht]
\centering
\begin{tikzpicture}[scale=1.2]

  \draw[-latex] (-0.5, 0) -- (6, 0) node (x) [right] {Space (x)};
  \draw[-latex] (0, -0.5) -- (0, 5.5) node (t) [above] {Time (t)};

  \foreach \i in {0,1,2,3} {
    \pgfmathsetmacro{\x}{\i + 0.5}
    \pgfmathsetmacro{\t}{\i + 0.5}

    \draw [fill=blue!20, opacity=0.6, draw=blue!80!black, thick]
      (\x, \t) ellipse (0.4 and 0.6);

    \filldraw [blue!60!black] (\x, \t) circle (2pt);
  }

  \draw [fill=red!30, opacity=0.6, rounded corners=3pt]
    (1.0,0.3) -- (3.0,2.3) -- (3.4,2.7) -- (1.4,0.7) -- cycle;

  \node [red!70!black, font=\small, anchor=west] at (4., 3.0) {QIX-C region};

  \draw[-latex, very thick, red!70!black, opacity=0.8] (0.6,0.3) -- (3.8,3.5) node [above right, align=left, font=\small] {Compression trajectory};

  \node [align=center, font=\small] at (4.7, 1.0) {Negative\\energy pulses};

\end{tikzpicture}
\caption{Schematic of a QIX-C formation from synchronized QET array gates. Each pulse generates a local negative energy region, which overlaps in time and space to form a traveling curvature compression packet.}
\label{fig:Fig12_QIXC}
\end{figure}
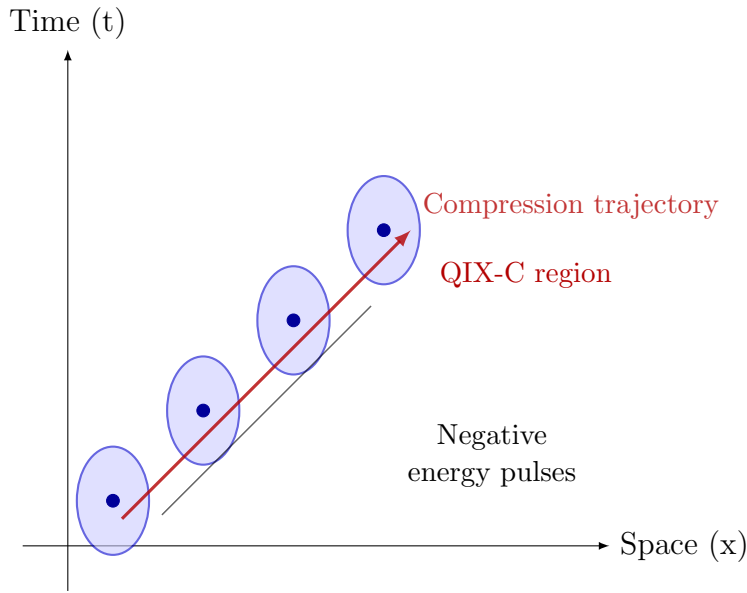


\begin{figure}[htbp]
    \centering
    \includegraphics[width=\linewidth]{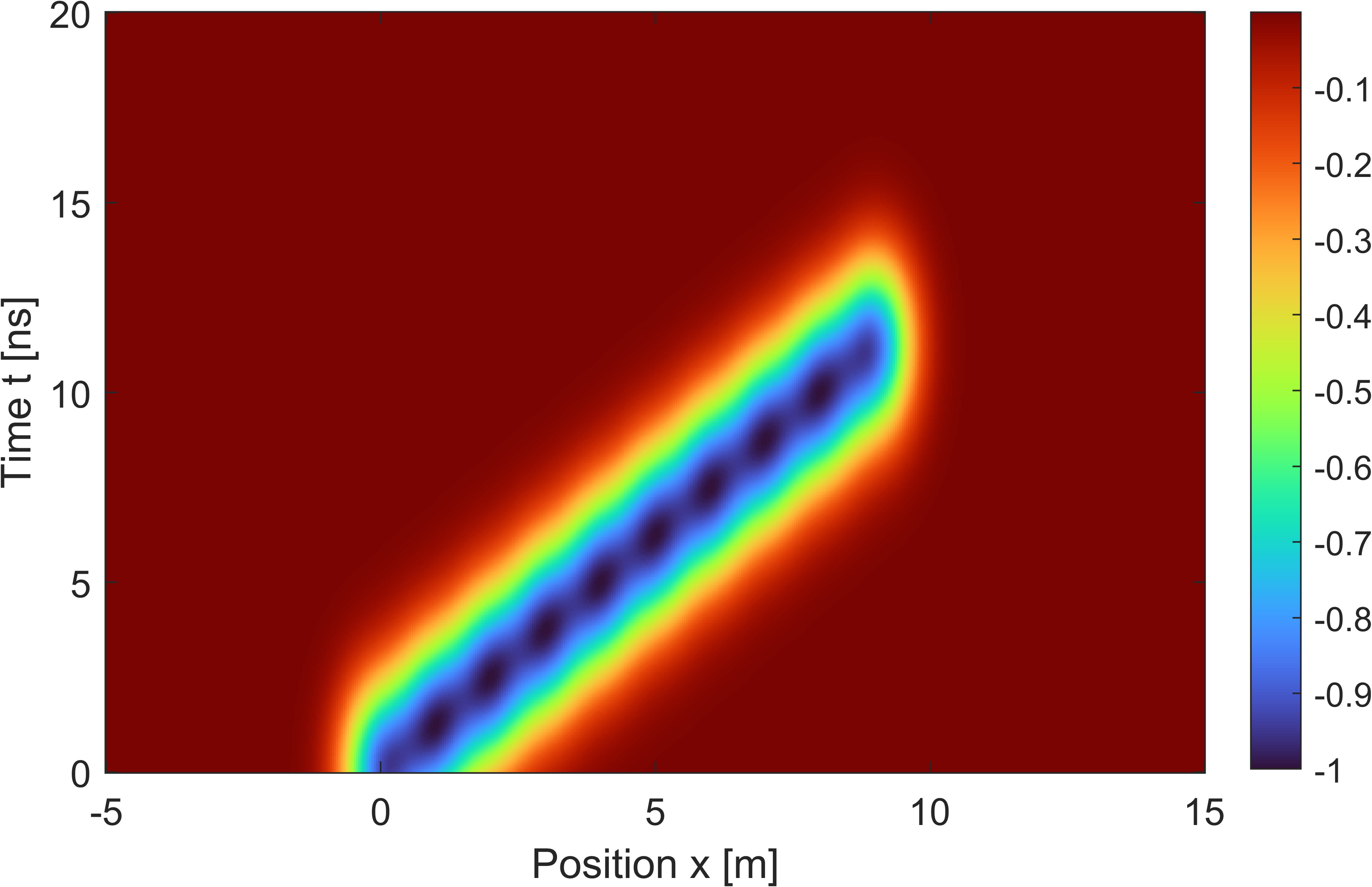}
    \caption{Simulated Ricci curvature profile \(\delta R(x,t)\) from a 1D QET array generating a traveling compression channel. The moving negative curvature dip results from constructive overlap of timed QET-induced stress–energy. Parameters: \(N = 10\), \(d = 1\), \(\sigma = 0.5\), gate interval = \(0.75\,\sigma/c\).}
    \label{fig:Fig13qixprofile}
\end{figure}
\DSZ 

\section{Conclusion and Outlook}
\label{sec:conc_9}

We have presented a detailed theoretical and simulation-based investigation of localized curvature effects generated by quantum energy teleportation (QET) protocols. Building on the ER=EPR paradigm and negative energy concentration techniques, we modeled how stress-energy imprints from QET sequences lead to transient curvature fluctuations that are, in principle, detectable using precision metrology platforms.

In Section~\ref{sec:FocNegEnergy_4}, we quantified the relationship between detector geometry, field correlations, and the resulting curvature signal $\delta R(x,t)$ using a Gaussian model for $\langle T_{00} \rangle$. We showed that squeezing, synchronization, and spatial shaping strategies can enhance signal amplitude while mitigating noise and decoherence. A parametric signal-to-noise model and its visualization (Fig.~\ref{fig:Fig2_snr_contour}) highlighted viable regimes for curvature detection under realistic constraints.

In Sections~\ref{sec:QIXC_8}, we introduced the concept of Quantum-Curvature Compression (QIX-C) as an extended application of synchronized QET arrays. These structures amplify curvature via pulse-shaping, temporal gating, and multimode enhancement. Simulated curvature profiles and clock drift signatures (Figs.~\ref{fig:Fig8_clock_drift_vs_deltaR}–\ref{fig:Fig10_clock_drift_combined}) demonstrated plausible observables for upcoming precision experiments.

\textbf{Outlook.} Experimental access to quantum-induced curvature lies at the intersection of quantum optics, relativistic field theory, and gravitational metrology. In the near term, laboratory-scale implementations using squeezed light, cavity-enhanced QET, and atomic clocks could test weak-field predictions presented here. Space-based variants may eventually allow for coherent arrays spanning long baselines, leveraging gravitational isolation and ultrastable clock networks.

Future work will extend the analysis to full quantum field backreaction, include stochastic fluctuations in the energy-momentum tensor, and explore holographic or conformal field theory mappings for more exotic energy configurations. Our results suggest that curvature induced by local operations on quantum fields — long thought to be inaccessible — may be testable with upcoming quantum technologies.

\section*{Acknowledgments}

The author gratefully acknowledges stimulating discussions with colleagues in the quantum foundations and quantum information communities, whose insights helped sharpen the conceptual framework of this work. Special thanks to the faculty and mentors at the Johns Hopkins University Advanced Academic Program for continued guidance and encouragement. Portions of this research benefited from open-access resources, including arXiv and the NASA ADS database. No external funding was used for this study.

This manuscript also benefited from the use of OpenAI’s ChatGPT (GPT-4), particularly for LaTeX structuring, technical editing, reference cleanup, and table formatting during manuscript preparation~\cite{chatgpt2025,openai2023}. Computations were performed using Computations and simulations were performed using MATLAB\textsuperscript{\textregistered} (MathWorks, Natick, MA) \cite{MATLAB2024}. All scientific content, interpretation, and conclusions are the responsibility of the author.

This work received no external funding. All opinions and conclusions are those of the author and do not represent institutional positions.

\vspace{1em}
\noindent


\appendix

\renewcommand{\thefigure}{\thesection\arabic{figure}}
\renewcommand{\thetable}{\thesection\arabic{table}}
\setcounter{figure}{0}
\setcounter{table}{0}

\renewcommand{\thesubsection}{\thesection.\arabic{subsection}}

\section{Detailed Modeling of Negative Energy-Induced Curvature and Signal-to-Noise Ratio}
\label{appendix:snr_model}

This appendix consolidates the key theoretical relations governing the generation of localized negative energy density via Quantum Energy Teleportation (QET), the induced spacetime curvature response, and the resulting signal-to-noise ratio (SNR) in gravitational detection experiments.

\subsection{Negative Energy Density Profile}

The QET protocol produces a localized negative energy density pulse approximately modeled as a Gaussian in space and time:
\begin{equation}
\langle T_{00}(x,t) \rangle \approx -\epsilon \exp\left[-\frac{(x - x_0)^2}{2\sigma^2} - \frac{(t - t_0)^2}{2\tau^2} \right],
\label{eq:neg_energy_gaussian}
\end{equation}
where \(\epsilon > 0\) denotes the magnitude, \(\sigma\) is the spatial width, and \(\tau\) the temporal width of the pulse centered at \((x_0, t_0)\).

\subsection{Curvature Response}

In the weak-field limit and harmonic gauge, the linearized Einstein equation relates the metric perturbation to the stress-energy tensor by:
\begin{equation}
\Box \bar{h}_{\mu\nu}(x) = -16 \pi G \langle T_{\mu\nu}(x) \rangle,
\label{eq:linearized_einstein}
\end{equation}
where \(\bar{h}_{\mu\nu} = h_{\mu\nu} - \frac{1}{2} \eta_{\mu\nu} h\).

The scalar curvature perturbation \(\delta R\) is primarily sourced by \(\langle T_{00} \rangle\), yielding:
\begin{equation}
\delta R(t,x) = 8\pi G \, \delta \langle T_{00}(t,x) \rangle,
\label{eq:ricci_curvature}
\end{equation}
assuming localized negative energy regions dominate curvature changes.

\subsection{Time-Resolved Curvature and Clock Drift}

For a negative energy pulse with characteristic amplitude \(\delta R_0\), spatial extent \(L\), and duration \(\Delta t\), the induced fractional clock drift between co-located clocks is approximately:
\begin{equation}
\frac{\Delta \tau}{\tau} \approx \frac{1}{12} \delta R_0 \cdot L^2 \cdot \Delta t,
\label{eq:clock_drift}
\end{equation}
providing a direct observable for gravitational effects of QET-induced negative energy.

\subsection{Parametric Model for Signal-to-Noise Ratio}

The cumulative curvature signal from an array of QET units is enhanced by several factors, including the number of units \(N\), detector spacing \(d\), cavity finesse \(F\), entanglement depth \(G_{\mathrm{ent}}\), spatial shaping \(G_{\mathrm{shape}}\), and multimode coupling \(G_{\mathrm{multi}}\). Noise is reduced by repetition rate \(f\), optical squeezing parameter \(r\), and environmental shielding \(G_{\mathrm{noise}}\). These effects combine approximately as:

\begin{equation}
\mathrm{SNR} \sim \left( \frac{N}{d^3} \cdot \frac{F}{\pi} \cdot G_{\mathrm{ent}} \cdot G_{\mathrm{shape}} \cdot G_{\mathrm{multi}} \right) \Bigg/ \left( \frac{1}{\sqrt{f}} \cdot e^{-r} \cdot G_{\mathrm{noise}} \right).
\label{eq:snr_model_app}
\end{equation}
\DSZ 

This formula captures the leading parametric dependencies and forms the basis for the numerical evaluations visualized in the main text (e.g., Fig.~\ref{fig:Fig2_snr_contour}).

\subsection{Additional Notes}

\DSZ 
\begin{itemize}
 \renewcommand\labelitemi{--}
 \item The scaling \(1/d^3\) reflects enhanced energy density confinement in narrower Casimir gaps or smaller detector spacing.
\item The finesse term \(F/\pi\) accounts for coherent resonant enhancement in optical cavities.
\item The exponential term \(e^{-r}\) models noise suppression via squeezing parameter \(r\).
\item This model assumes idealized coherence and neglects higher-order decoherence or technical noise effects, which require detailed system-specific analysis.
\end{itemize}

This appendix serves as a self-contained summary of the mathematical framework underpinning the QET-induced negative energy curvature and experimental detectability.

\DSZ 

\section{Modeling of Curvature Profiles from QET Architectures}
\label{appendix:curvature_profiles}

This appendix outlines the modeling assumptions used to simulate the curvature profiles \(\delta R(x)\) shown in Figure~\ref{fig:Fig5_clock_drift_vs_deltaR}, corresponding to various QET architectures (single pair, unsynchronized array, and synchronized array).

\subsection{Energy Density Model}

Each QET unit induces a localized energy density perturbation in the quantum field, modeled as a temporally bounded Gaussian negative pulse:

\begin{equation}
    \langle T_{00}(x) \rangle = -\epsilon \exp\left( -\frac{(x - x_0)^2}{2\sigma^2} \right),
    \label{eq:appendix_T00_gaussian}
\end{equation}

where:
\begin{itemize}
 \renewcommand\labelitemi{--}
    \item \(\epsilon > 0\) is the magnitude of the negative energy density,
    \item \(x_0\) is the spatial center of the energy dip,
    \item \(\sigma\) is the characteristic width (smearing scale) of the field–detector interaction.
\end{itemize}

\subsection{Curvature Response}

In the weak-field static regime, the linearized Einstein equation relates the scalar curvature to the energy density:
\begin{equation}
    \delta R(x) = -8\pi G \, \langle T_{00}(x) \rangle.
    \label{eq:appendix_deltaR}
\end{equation}

Using Eq.~\ref{eq:appendix_T00_gaussian}, we model the curvature profile as:

\begin{equation}
    \delta R(x) = 8\pi G \epsilon \exp\left( -\frac{(x - x_0)^2}{2\sigma^2} \right).
    \label{eq:appendix_deltaR_gaussian}
\end{equation}

\subsection{Array Architectures}

We simulate different QET configurations as superpositions of such Gaussian profiles:

\paragraph*{Single QET Pair:}
A solitary negative energy dip centered at \(x_0 = 0\) with smearing width \(\sigma = 0.1~\mathrm{m}\).

\paragraph*{Uncoordinated Array:}
Five QET pairs placed at random positions \(x_i\) with phase-incoherent coupling, modeled as a sum:
\[
\delta R(x) = \sum_{i=1}^5 8\pi G \epsilon_i \exp\left( -\frac{(x - x_i)^2}{2\sigma^2} \right),
\]
with identical amplitudes \(\epsilon_i\) but unaligned phases and spread positions.

\paragraph*{Synchronized Array:}
Five QET pairs are tightly phase-aligned and symmetrically spaced around the origin, producing constructive overlap:
\[
\delta R(x) = 8\pi G \epsilon \sum_{i=-2}^{2} \exp\left( -\frac{(x - i\Delta x)^2}{2\sigma^2} \right),
\]
where \(\Delta x \ll \sigma\) ensures spatial compression.

\DSZ 

\subsection{Normalization and Units}

We assume a fiducial energy scale \(\epsilon \sim 10^{-11}~\mathrm{J/m^3}\), yielding curvature amplitudes \(\delta R_{\mathrm{peak}} \sim 10^{-36}~\mathrm{m^{-2}}\). These are consistent with expected stress-energy magnitudes in sub-eV scale QET protocols.

\subsection{Numerical Plotting}

The curvature profiles in Figure~\ref{fig:Fig5_clock_drift_vs_deltaR} are generated using MATLAB code that evaluates Eq.~\ref{eq:appendix_deltaR_gaussian} (or its sums) on a spatial grid. The Gaussian widths \(\sigma\) are chosen to reflect experimentally plausible smearing scales between 0.05 and 0.2 meters.

\vspace{1.0cm}


\bibliographystyle{iopart-num}

\end{document}